# Towards a consistent understanding of the exotic nucleus $^{42}_{14}Si_{28}$

Syed Afsar Abbas[a], Anisul Ain Usmani[b], Usuf Rahaman[b*] & M Ikram[b]

[a]Centre for Theoretical Physics, J M I University, New Delhi 110 025, India
[b]Department of Physics, Aligarh Muslim University, Aligarh 202 002, India

The issue of magicity of $^{42}_{14}Si_{28}$ has been a contentious one. Fridmann *et al.*, through studies of two-proton knockout reaction $^{44}_{16}S_{28} \rightarrow {}^{42}_{14}Si_{28}$, presented strong evidence in support of magicity and sphericity of $^{42}_{14}Si_{28}$. However in complete conflict with this, Bastin *et al.*, gave equally strong empirical evidence, to show that the N = 28 magicity had completely collapsed in $^{42}_{14}Si_{28}$ to make it well deformed. The consensus at present though is in favour of the validity of the latter experiment. However, our QCD based theoretical model supports the result of Fridmann experiment. They had explored the amazing persistence of the unique exotic nucleus $^{42}_{14}Si_{28}$ as a stable structure within the nucleus $^{48}_{20}Ca_{28}$; even after stripping off six-protons through the isotonic chain: $^{48}_{20}Ca_{28} \rightarrow {}^{46}_{18}Ar_{28} \rightarrow {}^{44}_{16}S_{28} \rightarrow {}^{42}_{14}Si_{28}$. Thus it is the novel and unexpected stability of proton shell closure at Z=14 in $^{42}_{14}Si_{28}$, which is playing such a dominant role in ensuring its magicity, while the neutron magic number N = 28, goes into hiding or actually disappears. Recently, SAA has shown that the fusion experiment of a beam of halo nucleus $^6He$ with the target nucleus $^{238}U$, actually provided strong evidence that the "core" of the halo nucleus has the structure of a tennis-ball (bubble) like nucleus, with a "hole" at the centre of its density distribution. This provides us with clear-cut support for our Quantum Chromodynamics based model of clusters of tritons in neutron-rich nuclei. Here we show that our QCD based model, provides support to Fridmann *et al.*, showing that, $^{42}_{14}Si_{28}$ has a spherically magic structure of 14 $^3_1H_2$ (14-tritons) with a tennis-ball (bubble) like structure with "hidden" N=28 neutrons.

## 1 Introduction

The magicity imposed on a particular neutron or proton number doesn't appear as universal phenomena throughout the periodic table. Some time new magicity rises and often prominent magicities fail to show their impact. The issue of whether $^{42}_{14}Si_{28}$ is magical or not has been a contentious one. Fridmann *et al*[1,2], through studies of the two-proton knockout reaction $^{44}_{16}S_{28} \rightarrow {}^{42}_{14}Si_{28}$, presented strong empirical evidence in support of magicity and sphericity of $^{42}_{14}Si_{28}$. However, Bastin *et al.*[3], gave equally strong evidence, but based on different empirical information, to show that the N=28 magicity had completely collapsed. Gade *et al.*[4] have also confirmed the breakdown of the N=28 magic number in $^{42}_{14}Si_{28}$.

Now let us revisit the experimental result of Fridmann *et al*.[1,2]. What they essentially explored was the amazing persistence of the unique exotic nucleus $^{42}_{14}Si_{28}$ as a stable structure within the nucleus $^{48}_{20}Ca_{28}$; even after stripping off six-protons through the isotonic chain: $^{48}_{20}Ca_{28} \rightarrow {}^{46}_{18}Ar_{28} \rightarrow {}^{44}_{16}S_{28} \rightarrow {}^{42}_{14}Si_{28}$. Thus it is the novel stability of proton shell closure at Z=14 in $^{42}_{14}Si_{28}$, which is playing such a dominant role in ensuring its double magicity within $^{48}_{20}Ca_{28}$. Thus the dominant role of magicity of Z=14 was basic. As such this experiment[1] was not making any direct statement about the magicity of the corresponding neutron number at N=28. However one had to make an extra assumption of the independent existence of a stable neutron structure at N=28, to be able to treat this nucleus as being doubly magical. They showed[2] that reducing the shell gap for N=28 did not affect the two-proton knockout cross section. Note that the insensitivity of N=28 magic number to the stability and sphericity imposed at proton number Z=14, is a completely new and unexpected reality of the structure of $^{42}_{14}Si_{28}$. It means the neutron magic number N=28 has actually become inoperative, or that it has gone into hiding here.

Now in as much as what the two-proton knock out reaction cross section, as studied by Fridmann *et al*[1,2], leads to the above clear and direct conclusion; and which is that this strong shell closure of proton number at Z=14 is so dominant that it leads to extra stability, magicity and sphericity of $^{42}_{14}Si_{28}$, and that the same is independent of the neutron magic number. Thus what Fridmann *et al.* have found is a new and novel structure of the exotic nucleus $^{42}_{14}Si_{28}$, and which goes beyond our conventional understanding of nuclear structure. But this novel property of $^{42}_{14}Si_{28}$ has been missed so far, mainly due to the dominating influence of the assumption that proton and neutron were the only degrees of freedom even in the exotic nuclei. The fact that simultaneously there was another experiment[3], that showed the same nucleus as displaying strong deformation at N=28 through the study of a low lying $2^+$ state, added to the confusion. Jurado *et al.*[5] didn't observe shell structures to change around N=28 for Si unlike P and S. They have interpreted it as the perseverance of N=28 shell closure or sudden change in the deformation in $^{42}_{14}Si_{28}$. Thus the two options may actually coexist simultaneously to provide the essential duality here.

---

*Corresponding author (E-mail: urahaman@myamu.ac.in)

## 2 QCD Based Model

Recently, one of the authors (SAA) has shown[6] that the fusion experiment of an incoming beam of halo nucleus $^6$He with the target nucleus $^{238}$U, actually provided strong and unambiguous evidence that the structures of the target nucleus (having standard nuclear density distribution described with canonical RMS radius r = $r_0 A^{1/3}$ with $r_0$=1.2 fm) was completely different from that of the "core" of the halo nucleus, which does not follow the standard density distribution with the above RMS radius. In fact, the core has the structure of a tennis-ball (bubble) like nucleus, with a "hole" at the centre of the density distribution. This provides us with clear-cut support for our model of the halo nucleus[7]. One point we would like to emphasize here - that right from the first proposal of the QCD based model in 2001[7], SAA had made unique prediction that the nucleus $^{42}_{14}Si_{28}$, is a clear tennis-ball (bubble) like nucleus with a hole at the centre of its density distribution.

The Fermi distribution matches the nuclear density distribution:

$$\rho = \frac{\rho_0}{1 + \exp\left(\frac{r-c}{a}\right)}$$

Here parameter c is defined as where the density comes down to $\frac{\rho_0}{2}$, with $\rho_0$ as the density at the centre; the surface thickness parameter s = 4.40a ~ 2.40 fm. This standard nuclear density distribution is described by the canonical RMS radius r = $r_0 A^{1/3}$ with $r_0$ = 1.2 fm.

The density of the above target nucleus is clearly given by the above Fermi distribution. This is shown typically like that of say, bismuth in Fig. 1. But as per the conclusion of paper [6], the core of the halo-nucleus density distribution is clearly unlike it, and this has a hole at the centre, as shown schematically in the inset of Fig. 1. So the core of the halo-density density distribution is fundamentally different from that of the standard target nucleus, what degree of freedom may explain this? In the paper[6], it was shown that this new degree of freedom was the triton. The neutron-rich core nuclei $^{3Z}_Z X_{2Z}$, are made up of Z $^3_1H_2$ clusters, and these created the tennis-ball like structure as shown in the inset of Fig. 1. It is not made of simple proton and neutrons, but of clusters of tritons, treated as elementary entities.

## 3 Triton Clustering in Nuclei

Now we will attempt to legitimize our claim to have a group of tritons inside a N = 2Z neutron-rich system. We look for evidences where the triton has appeared as an elementary entity. In reference[7] a new group $SU_\mathcal{A}(2)$ termed as nusospin has been introduced[7]. It is similar to the SU(2) group but difference is $SU_\mathcal{A}(2)$ treats the pair (h,t) as fundamental representation in place of (p,n). The physical defense of this new model is likewise examined in detail. On the side of the nusospin group, we have discovered solid confirmations preferring A = 3 clustering in nuclei as appeared in reference[8]. We there, similarly as light N = Z nuclei with A = 4n, n = 1, 2, 3, 4 . . . might be treated as being made out of n-α cluster[9], we have shown earlier in our paper[8], a few neutron-rich cores which might be treated as being made out of n-clusters of $^3_1H_2$. The binding energy light neutron-rich nuclei as $E_B$ = 8.48n + Ck, where 8.48 MeV is the binding energy of $^3_1H_2$. We assume these n-clusters of $^3_1H_2$ to form k bonds with C inter-triton-bond energy. We have considered the exact same geometric structure of clusters in these nuclei as conventionally done for α- clusters in A = 4n nuclei. Thus, the model seems to hold out well with inter-triton cluster bond energy of about 5.3 MeV. In reference[7], the author argued that the estimation of inter-triton-bond energy even holds for heavier nuclei like $^{42}_{14}Si_{28}$.

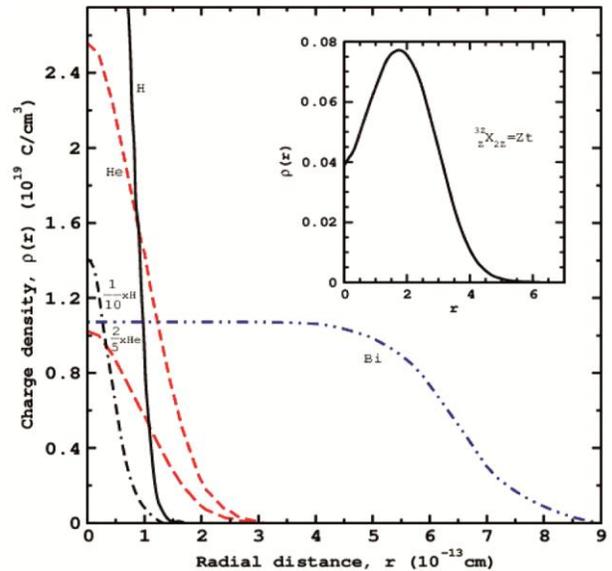

Fig. 1 — Schematic density distribution of nuclei as determined by electron scattering. Inset (t for triton) shows the same with a marked "hole" at the centre as that of the core of the halo nucleus. and what is called a tennis-ball (bubble) like structure. Note the basic difference between the structures of t and Bi.

Next, we call attention to experimental proof of the conceivable presence of helion and triton clusters in $^6$Li cores. In reality, the equivalent has been convincingly shown through direct trinucleon knock out, from $^6$Li exclusively by means of electron reaction[12]. The momentum transfer dependence estimated in two mirror reactions $^6$Li(e, e'$^3$He)$^3$H and $^6$Li(e, e'$^3$H)$^3$He were seen as in complete conflict with the basic spectrum of a direct single nucleon knockout reaction. Then again, the momentum transfer was in acceptable concurrence with a direct A = 3 knockout reaction. This obviously shown h-and t-clusters existed as essential elements in $^6$Li.

In analogy with the fact that we know as per mean-field concept, that a bunch of protons and neutrons in a nucleus, would create an average binding potential for each nucleon, we assume that a bunch of elementary tritons in a nucleus too would create an average binding potential for each triton in a nucleus. It is such a potential, which is binding tritons in these neutron-rich nuclei with $^{3Z}_ZX_{2Z} = Z^3_1H_2$; that is, these nuclei are made up of Z number of tritons. Thus we extract one-triton separation energies of these pure triton constituent nuclei. Let us define $S_{1t}$ = $B(^{3Z}_ZX_{2Z}) - B(^{3Z-3}_{Z-1}Y_{2Z-2}) - B(^3_1H_2)$ where, $B(^A_ZX_N)$ is the binding energy of the nucleus $^A_ZX_N$.

## 4 Magicity of $^{42}_{14}Si_{28}$

However as our focus in this paper is the issue of the magicity of $^{42}_{14}Si_{28}$, we concentrate on the study of nuclei in the vicinity of this nucleus. In Fig. 2 we display our RMF result with NL3 interaction along with the presently available experimental data[11] also. Note the clear RMF model prediction of magicity of $^{24}_8O_{16}$ and $^{60}_{20}Ca_{40}$. However, on closer scrutiny of the structure between the two extremes of the strongly magical nuclei $^{24}_8O_{16}$ and $^{60}_{20}Ca_{40}$, we notice a prominent broad hump or "plateau of stability". We may treat this hump as a broad "peak" of stability, and take it as all those being magical, and so justifiably call it a "plateau of magicity". This plateau of magicity is being defined by the two boundary towering peaks of magicity at $N_t$ = 8 ($^{24}_8O_{16}$) and $N_t$ = 20 ($^{60}_{20}Ca_{40}$), respectively. However equally significant, in defining this plateau of magicity, are the two boundary nuclei manifesting themselves as extremely deep trenches at $N_t$ = 9 ($^{27}_9F_{18}$) and $N_t$ = 21 ($^{63}_{21}Sc_{42}$).

In Fig. 2, there appears, a slight kink, at $^{42}_{14}Si_{28}$, and which is somewhat more stable than the nuclei surrounding it, viz $^{36}_{12}Mg_{24}$ and $^{48}_{16}S_{32}$. This is also placed at the centre of its plateau of magicity. Thus $^{42}_{14}Si_{28}$ should be considered as more of a doubly magic nucleus than the other members of the plateau of magicity. It has been a long standing paradigm in nuclear physics that the central potential is proportional to the ground state baryon density and a spin-orbit potential proportional to the derivative of the same central potential. Remarkably Todd-Rudel[12] found that the dramatic decrease in spin-orbit splitting as seen in exotic nuclei is not caused by the neutron density in the nuclear surface but by proton density in the nuclear interior. In that paper [12] they found within RMF model calculations with NL3 interaction, that as two-protons are removed from $^{48}_{20}Ca_{28} \rightarrow {}^{46}_{18}Ar_{28}$, the standard density of $^{48}_{20}Ca_{28}$ (e.g. as in Fig. 1 for nuclei like Bismuth) quickly transforms into a hole-like nucleus for $^{46}_{18}Ar_{28}$ itself. But this fails to reproduce the basic putative property of the amazing persistence of the nucleus $^{42}_{14}Si_{28}$ as a stable structure within the nucleus $^{48}_{20}Ca_{28}$. What is the reason for the RMF model with NL3, to have failed to reproduce this essential property of $^{42}_{14}Si_{28}$. Piekarewicz realized [13] that this had to do with the fact that the NL3 interaction was failing to produce the $1d_{3/2}$ - $2s_{1/2}$ proton gap in $^{40}Ca$, in the first place. It gave a proton gap of only 0.83 MeV, while experimentally it was about 2.8 MeV. So he tweaked the NL3 parameters slightly, in a minimal manner, so that this basic problem of the Calcium-chain was rectified. Right away he could get consistent point proton density distribution of all the nuclei in the basic six-protons stripping isotonic chain: $^{48}_{20}Ca_{28} \rightarrow {}^{46}_{18}Ar_{28} \rightarrow {}^{44}_{16}S_{28} \rightarrow {}^{42}_{14}Si_{28}$. We reproduce his Fig. 4 [13], as our Fig. 3 here.

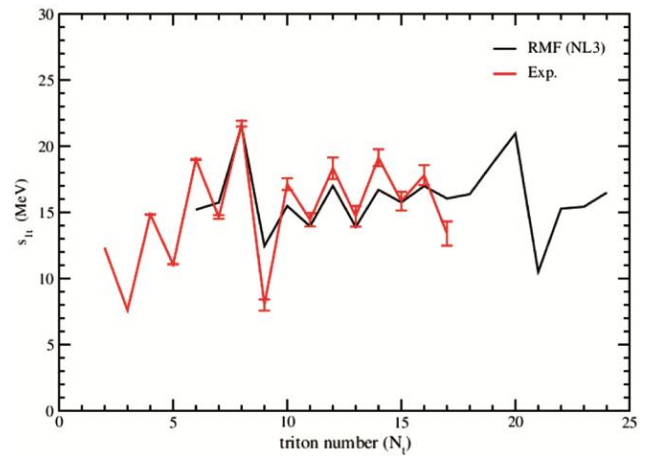

Fig. 2 — Triton separation energy [10]

In the inset, we show how this is almost equivalent to stripping six-protons from $^{48}_{20}Ca_{28}$ itself. Most remarkably Piekarewicz was thus able to explain physically as to what was happening in the experiment by Fridmann *et al.*[1]. First the study of proton single particle spectrum of RMF model calculations in the chain: $^{40}_{20}Ca_{20} \rightarrow {}^{48}_{20}Ca_{28} \rightarrow {}^{42}_{14}Si_{28}$, showed near degeneracy of proton orbital $1d_{3/2}$ - $2s_{1/2}$ in $^{48}_{20}Ca_{28}$, and the emergence of a strong Z=14 gap in $^{48}_{20}Ca_{28}$, and which persisted robustly in $^{42}_{14}Si_{28}$ [please see his [13] Fig. 1]. Next, the most amazing was how the neutron single particle spectrum behaved. Best to quote him[13], "Yet the present relativistic mean-field model predicts that as protons are progressively removed from the $1d_{3/2}$ - $2s_{1/2}$ orbitals, $1f_{7/2}$ neutron orbit returns to its parent fp-shell-leading to the disappearance of the magic number N=28. Thus in the present model, the proton removal is ultimately responsible for the return of the $1f_{7/2}$ neutron orbit to its parent shell". This disappearance of the N=28 magic number is exactly what Fridmann *et al.* had extracted experimentally [1,2] as we had discussed above. We have seen how Piekarewicz's paper[13] is able to explain and justify the empirical conclusions of Fridmann *et al.* work [1,2].

So far we have been used to talking of sphericity and magicity when both the proton and neutron numbers are separately and simultaneously magical. However here we are being compelled by the empirical reality, to talk of sphericity and magicity of $^{42}_{14}Si_{28}$ where only proton number Z=14 shell closure is playing a role, while the corresponding neutron number magic number N=28 has disappeared and gone into hiding. This demands an understanding within our theoretical picture of nuclear physics where we treat $^{42}_{14}Si_{28}$ nucleus to make up of fourteen-tritons. Thus 14-tritons are a bound state in a potential binding these tritons as elementary entities. This nucleus is an extra-bound state as it is closing the triton-shell orbital $d_{5/2}$ at triton-number $N_t$= 14. This is the same as proton number Z=14, and thus this is what is seen in our shell model analysis. As to neutrons, however, as each triton has two neutrons hidden inside a triton (similar to the way that 2-u and 1-d quarks are hidden inside a proton within a nucleus), in all 28-neutrons are hidden inside the 14-tritons in this magical and spherical tritonic nucleus $^{42}_{14}Si_{28}$. Thus physically relevant is only one magical number $N_t = 14 \sim Z = 14$.

## 5 Conclusions

In summary, $^{42}_{14}Si_{28}$ is made up of $N_t = 14$ number of tritons. This is the same as the number of protons making up this exotic nucleus. This one degree of freedom triton-shell model needs this triton number to close the $d_{5/2}$ orbital. The neutrons here are hidden inside these 14-tritons and thus physically they go out of contention in this case. So we may actually treat these 14-tritons as 14-quasi-protons, with the same charge as protons but each being much heavier due to the two neutrons hidden within its guts. Thus $^{42}_{14}$Si$_{28}$ is magical and spherical too. Most significantly, it has a hole at the centre of its density distribution. This is exactly what Fridmann *et al.*[1,2] have found experimentally.

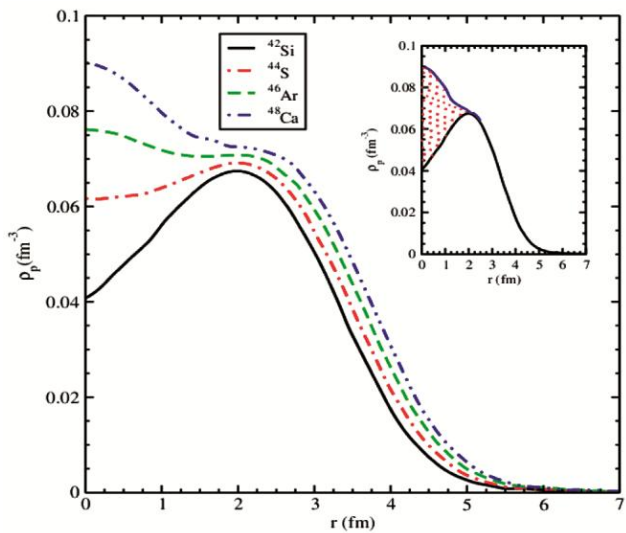

Fig. 3 — Point proton density of N=28 isotonic chain - schematic plot from Fig. 4 of [6].

**Acknowledgement**

Usuf Rahaman would like to thank University Grant Commission (UGC) for providing UGC-SRF fellowship. M Ikram would like to acknowledge the financial support in the form of Research Associateship by CSIR, New Delhi.